\documentclass[prd,reprint,longbibliography,amsmath]{revtex4-1}

\usepackage{graphicx}
\usepackage{hyperref}

\newcommand{\beq}{\begin{eqnarray}}
\newcommand{\eeq}{\end{eqnarray}}

\begin{document}

\title{%
  Inability to find justification of a $k_T$-factorization formula by following
  chains of citations
}

\author{Emil Avsar}
\email{eavsar@phys.psu.edu}
\author{John C. Collins}
\email{collins@phys.psu.edu}
\affiliation{%
  104 Davey Lab, Penn State University, University Park, 16802 PA, USA
}

\begin{abstract}
  Fundamental to much work in small-$x$ QCD is a $k_T$-factorization
  formula.  Normal expectations in theoretical physics are that when
  such a result is used, citations should be given to where the
  formula is justified.  We demonstrate by examining the chains of
  citations back from current work that violations of this expectation
  are widespread, to the extent that following the citation chains, we
  do not find a proof or other justification of
  the formula.  This shows a substantial 
  deficit in the reproducibility of a phenomenologically important area
  of research.  Since the published formulae differ in normalization,
  we test them by making a derivation in a simple model that obeys the
  assumptions that are stated in the literature to be the basis of
  $k_T$-factorization in the small-$x$ regime.  We find that we disagree
  with two of the standard normalizations.
\end{abstract}

\maketitle

\section{Introduction}

Recently, the participants in a Les Houches meeting published a set of
recommendations for the presentation by experimental groups of the
results of searches for new physics at the LHC \cite{Kraml:2012sg}.
The recommendations are based on the following guiding principles:
\begin{enumerate}
\item[(E1)] What has been observed should be clear to a non-collaboration colleague.
\item[(E2)] How it has been observed should be clear to a non-collaboration colleague.
\item[(E3)] An interested non-collaboration colleague should be able to use and (re)-interpret results
without the need to take up the time of collaboration insiders.
\end{enumerate}
(The labeling is ours.)

In this paper we will argue that principles of a similar kind should
be applied to theoretical papers, and that it is a substantial
impediment to progress in high-energy physics that these principles are
not met by a large number of published papers.  We will support our
argument with examples from the literature on the Regge, or small-$x$,
limit of QCD.  This area is highly relevant for the LHC, thereby
giving a natural connection with the recommendations for experimental
results.

The processes of theoretical physics are rather different to those of
experimental physics, so a direct transcription of the principles for
presenting experimental results is inappropriate.  Instead we propose
the following:
\begin{enumerate}
\item[(T1)] What has been calculated or derived should be clear to any
  colleague. This has a consequence that for the concepts involved,
  either precise definitions should be provided or correct references
  to where correct definitions can be found.

\item[(T2)] It should be clear to any colleague whether the obtained results are at the level 
of a mathematical proof, a conjecture, a guess etc. 
 
\item[(T3)] Any interested colleague, with appropriate expertise and
  knowledge of the underlying theory, should be able to understand,
  use, and, most importantly, reproduce the results without the need
  to obtain extra explanation from the author(s).  

\end{enumerate}

Now it is fundamental to science that results should be reproducible.  In
theoretical physics this means that an interested reader can, for
example, replicate the derivations and calculations in a scientific
article.  It is important to be able to do this independently of the
author(s) of the original article.  The derivations in a specific
paper are not always self-contained, and references to 
where proper derivations can be found are then essential.  An exception
is when adequate derivations are genuinely part of the expected
education or experience of the intended audience of an article.
So a corollary to (T3) is what should be standard practice:
\begin{enumerate}
\item[(T4)] When previous results are relied on, appropriate
  references must be made to where appropriate justifications are
  actually to be found. This need not apply if the results are a
  standard part of the education of scientists working in the general
  area (e.g., theory of elementary particles).
\end{enumerate}
A general application of the
principles (T1)--(T4) is necessary for the progress of our field.  We
consider them to represent the traditional standards of good
theoretical physics.

For reproducibility of results, there is an important difference
compared with experimental physics.  Replicating an experimental
result can often mean an amount of work comparable to that for the
initial result, although, as time goes by, initially highly difficult
techniques can become routine.  (E.g., the measurement of jets in
high-energy collisions.)  In theoretical physics, the initial
formulation and derivation of particular results may require a high
degree of unusual creativity and insight.  But the verification of the
results should be much more routine.

It is not uncommon that a theoretical argument involves unstated
assumptions or tacit knowledge that is not explicitly codified.  In
such cases extensive consultation with the authors of an article might
be needed before the results of the article can be reproduced.  If
widespread, such a
situation is highly undesirable and is in the long run inimical to
good science.  

Some of the tacit knowledge may not be easy to formulate explicitly,
and has to be acquired by appropriate guided exposition.  We consider
classic examples to be the concepts of mass and force in Newtonian
mechanics, which resist a precise definition in terms of pre-existing
concepts.  But for the health of a science, this kind of tacit
knowledge should be strongly limited in extent.

To counterbalance the negative consequences of our finding that some
results are not easily reproducible, it is important
to remember that subjects such as those we discuss are fundamentally
difficult, both conceptually and technically.  It is perhaps
inevitable that progress involves intuitive ideas acquired through
long experience working in an area.  Furthermore, it is genuinely
difficult to 
convert such intuitive ideas to a fully consistent and self-contained
mathematical structure.  

One motivation for the present paper came from an examination of the
literature on the Regge limit and the related small-$x$ limit of QCD.
These limits are widely applied to phenomenological studies and
predictions at HERA, RHIC and the LHC. We noted that the concept of
transverse-momentum-dependent (TMD), or $k_\perp$, factorization appears widely, and it is desirable to
unify or relate the various definitions of TMD distributions,
including those in the recent book by one of us \cite[Chs.\ 13 \&
14]{qcdbook}.  To do 
this systematically requires that we know sufficiently accurately what
the definitions are and how results for cross sections are obtained.
But in trying to do this, we discovered that in many cases,
comprehensible, accurate, 
and valid definitions and, proofs or other justification, are hard or
impossible to find, certainly if one just follows the citations given.
These problems have some resemblance to the 
issues found on the experimental side in Ref.\ \cite{Kraml:2012sg}, and has led 
us to formulate (T1)--(T4) above, as
reasonably self-evident principles of good practice in theoretical science.

The related topic of the QCD formulation of the Regge theory is
particularly interesting for our purposes.  Prior to the discovery of
QCD in 1972 to 1974, Regge theory was a primary topic of work in
the strong interaction.  Therefore there exists an older
generation of physicists who are/were knowledgeable in this subject, in particular 
in the fundamentals.  But with
the advent of QCD, work naturally strongly turned away from Regge
theory in favor of perturbative QCD.  (An INSPIRE search \texttt{find
  title Regge and date 1972} gave us 214 results, but \texttt{find title
  Regge and date 1983} gave 18.)

Even so, the phenomena addressed by Regge theory have not disappeared, and
small-$x$ physics makes heavy use of it (interestingly, the search \texttt{find
  title Regge and date 2010} gave 59 hits, showing an increase of interest in the old ideas). 
 Therefore for any young
physicist entering this field, there is a lot of knowledge that must
be learned from papers that deal with rather old concepts whose
foundation may not be particularly well-known today, even though the
concepts are widely used.  We are particularly concerned with
foundational issues, such as: Why are the concepts and results in the
subject what they are?  How are they defined?  How do we know they are
valid?  It is then necessary to go back to original papers, or even
simpler, textbooks on the subject (when available).  
Even with textbooks and review articles, principles (T1)--(T4) should
still apply.\footnote{Mark Strikman in a personal communication
  remarked that it is an interesting question as to whether the
  literature on Regge theory meets these standards sufficiently well.
  But we do not wish to answer that particular question here.}

The issue at hand is not just that one should make it easier for
newcomers to properly learn the field, but that it is also necessary
to be able to verify claims of theoretical results made in the
literature. 

Rather than give an exhaustive list of the cases where we have found
substantial difficulty verifying results stated in the literature, we
will take one particularly important case, $k_T$-factorization as
applied to the production of hadrons in hadron-hadron collisions.  We will
attempt to determine from some important papers on the subject how
this particular form of factorization is justified.  

We emphasize that it is \emph{not} necessary that there be an actual
rigorous proof of a formula that is used.  There are many situations
in theoretical physics where it is useful to propose or conjecture a
property or formula on the basis of some intuitive idea, for example,
or from some natural extension of existing ideas to new situations.
Even in these situations it is important to know what is the
justification for a formula, and why a particular formula is used
rather than one of the infinitely many other possible formulas.
Therefore we prefer to use the word ``justification'' rather than
``proof''.  Without knowing the status of the justification or proof
and without being able to evaluate the arguments, it is difficult to
evaluate the significance of subsequent work where problems are
encountered.

\section{An examination of the literature:
   Where is the origin of $k_T$-factorization?}

In this section we present explicit examples from the literature where we have
identified problems with the application of the principles (T1)--(T4),
in the case of
$k_T$-factorization in single inclusive particle
production in the small-$x$ regime.  The reasons for choosing this
particular subject are: 
(i) $k_T$-factorization has wide
phenomenological applications. (ii) It plays a very fundamental role
in the formulation of small-$x$ QCD. (iii) The concept has been around
a rather long time and therefore it is a reasonable expectation that
the subject is well established and that our principles are obeyed, so
that we can check the results.
Any problems in formulating the theoretical methods can be treated as
an important source of systematic error in the comparison of theory
and experiment.  
Explicitly noticing such problems can be an important
motivation for topics for further research that are important for the
success of a field.

The $k_T$-factorization formula that we discuss, Eq.\
\eqref{ktfact} below, is intended to be valid for single inclusive jet
production in hadron-hadron collisions.  It is widely used
in phenomenological applications to study the particle multiplicity
observed at hadron colliders.  (For some examples see
\cite{Kharzeev:2003wz, Armesto:2004ud, Kharzeev:2004if, Levin:2010dw,
  Levin:2010zy, Albacete:2010bs, Albacete:2010ad, Rezaeian:2011ia,
  Tribedy:2011aa} and references therein.  A comparison
of some phenomenological predictions to LHC data for the particle
multiplicities in both proton-proton and lead-lead collisions was
presented by the ALICE collaboration \cite{Aamodt:2010pb}.)

The $k_T$-factorization formula used in this area is (see, e.g., Ref.\
\cite[Eq.\ (1)]{Levin:2010zy}):
\begin{multline}
  \frac{d\sigma}{d^2p_T dy} 
  = \frac{2\alpha_s}{C_F \, p_T^2} 
\times \\ \times 
    \int d^2k_{A,T} \, f_A(x_A,k_{A,T}) \, f_B(x_B, p_T - k_{A,T}).
\label{ktfact}
\end{multline}
Here $f_A$ and $f_B$ are TMD densities of gluons in their parent
hadrons, and the two gluons combine to give an outgoing gluon of
transverse momentum $p_T$ which gives rise to an observed jet in the
final state.  In the formula, $C_F=(N_c^2-1)/2N_c$ with $N_c=3$ for
QCD, $y$ is the rapidity of the final-state jet, and $x_{A,B}=
(p_T/\sqrt{s}) e^{\pm y}$.  The incoming hadrons $A$ and $B$ can be
protons or nuclei.

Questions that now naturally arise are: Where does this formula
originate from?  Where can a proper derivation be found, and under
what conditions and to what accuracy is the derivation valid?  What
are the explicit definitions of the unintegrated distributions
$f_{A,B}$, and do these definitions overcome the subtleties that are
found in constructing definitions of TMD distributions in QCD in the
non-small-$x$ regime \cite[Chs.\ 13 \& 14]{qcdbook}?  In an ideal
world, we could say that in order for the requirements (T1)--(T4) to
be fulfilled, it is absolutely necessary that these questions be
answered, and that a person who reads a paper which makes use of this
formula can, if needed, go back to the original source and
himself/herself reproduce and verify the derivations.  But we must
recognize that in the real world some of these issues are very deep
and difficult, and that therefore complete answers to the questions do
not (yet) all exist.  Nevertheless, in this subject, we should expect
some kind of derivation, with the accompanying possibility of an
outsider being able to identify, for example, possible gaps in the logic where
further work is needed.

\begin{figure*}
\begin{center}
\includegraphics[angle=0, scale=0.6]{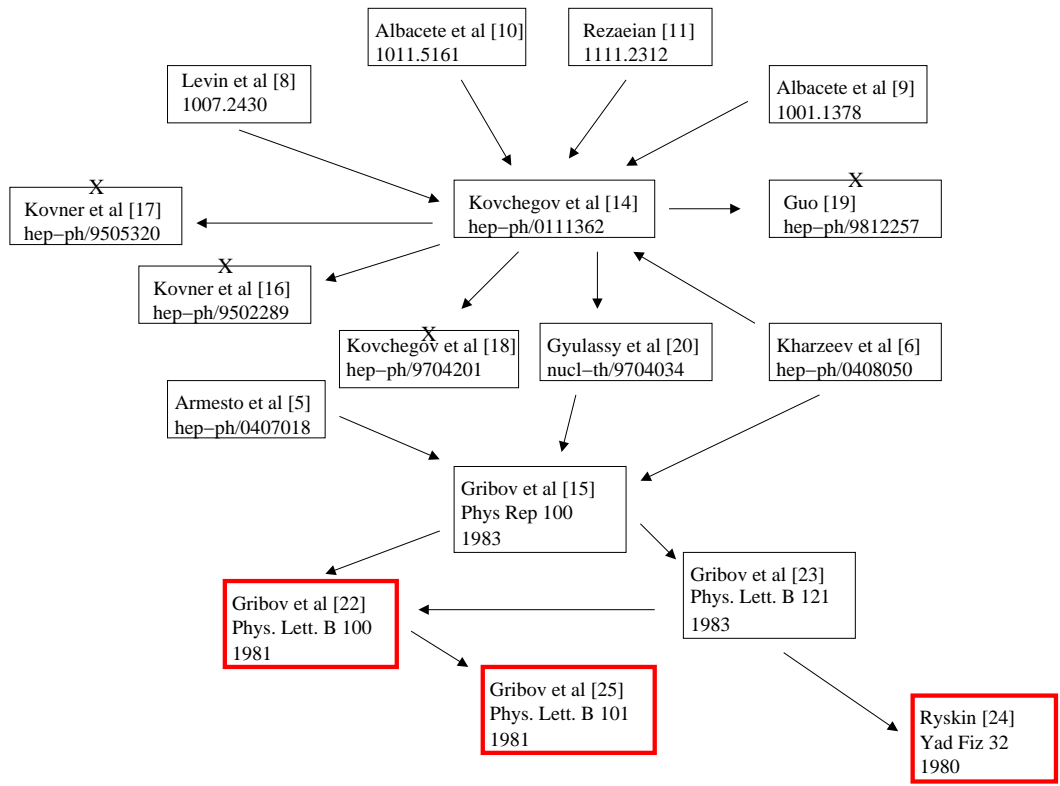}
\end{center}
\caption{\label{citationgraph} Some examples of the chain of
  references for the $k_T$-factorization formula \protect
  \eqref{ktfact}. The boxes with crosses on top indicate references
  quoted for \protect \eqref{ktfact} where the equation is not even
  present in the paper. The boxes surrounded by a thick red line at
  the end of a chain are where a final proof or adequate justification of the formula would be
  expected to be found, yet, as documented in the text, this is not the
  case. 
  Thus
  the chain of references does not lead to any source from which the
  validity of the formula could ultimately be verified.  Only the arXiv numbers are shown for those 
  papers that possess one, while for those who do not have arXiv numbers we show the journal name and volume
  for the publication. }
\end{figure*}

However, as we will explain below, we tried to find any kind of a 
derivation of the formula by following citations given for it,
but were unable to do so. 
Our findings can be visualized in Fig.\ \ref{citationgraph}, which
shows the chain of references that
one needs to follow to arrive 
at the nearest possible source(s), starting from a selection of
recent papers.  Coming to those sources we find
that the formula is never derived but essentially asserted. Moreover, the
basic concepts involved are never defined in a clear enough way to
make it understandable what exactly it is that is being done.
We
therefore find it impossible that \eqref{ktfact} can be satisfactorily
re-derived from sources referenced in the literature, contrary to what should
be the case if principles (T1)--(T4) hold.

A clear symptom that these are not merely abstract difficulties but
are problems with practical impact is that
the overall normalization factor differs dramatically
between the references.
See, for example, Eq.\ (40) in \cite{Kovchegov:2001sc}
and Eq.\ (4.3) in \cite{Gribov:1984tu} --- and notice that this
difference in normalization does not appear to be commented on, let
alone explained.  The difference in normalization factors demonstrates
that at least one of the presented factorization formulas is
definitively wrong.  (In the two appendices of the present paper, we
will show that in fact the normalizations of both formulas appear to
be wrong.)  There are a number of difficult physical and
mathematical issues that need to be addressed if one is to provide a
fully satisfactory proof of a factorization formula.  These issues go
far beyond a mere normalization factor. But the existence of problems
with the normalization factor is a diagnostic: it provides a clear and
easily verifiable symptom that something has gone wrong.  A minimum
criterion for a satisfactory derivation is that it should be explicit
enough to allow us to debug how the normalization factor arises.

At the top of our chart of references, Fig.\ \ref{citationgraph}, we have chosen 
some of the recent phenomenological applications \cite{Levin:2010zy, Albacete:2010bs, Albacete:2010ad, Rezaeian:2011ia} that make use of
\eqref{ktfact}. 
We also include some 
earlier highly cited phenomenological applications
\cite{Armesto:2004ud,Kharzeev:2004if}.  There exist a very great
number of papers which make use of \eqref{ktfact}, so 
we include here only a representative few. 
As is indicated in the top part
of Fig.\ 
\ref{citationgraph}, a central source that is given for \eqref{ktfact} is the highly
cited Ref.\ \cite{Kovchegov:2001sc}. 
We thus ask whether we then can find a derivation of \eqref{ktfact} in
\cite{Kovchegov:2001sc}. 

That paper performs a calculation in a quasi-classical approximation 
of particle production in DIS using the dipole formalism (see the reference for the exact 
calculations that define this ``quasi-classical'' approximation).  There actually is an 
implicit assumption of a factorized structure from the very start in this formalism (see
Eqs.\ (1) and (7) in the reference).
For our purposes it is important to notice what the exact statement 
is regarding \eqref{ktfact}, which can be found as
Eq.\  
(40) in \cite{Kovchegov:2001sc}.  (An unimportant difference is that in \cite{Kovchegov:2001sc}, the $f$'s in 
\eqref{ktfact} are instead written as  $f/k_T^2$.)  Prior to this
equation, an equation for the production of gluons 
in DIS is derived, Eq.\ (39) in \cite{Kovchegov:2001sc}. 
The exact statement just prior to stating \eqref{ktfact} 
in the form of Eq.\ (40) in \cite{Kovchegov:2001sc} reads
\begin{quote}
  \emph{The form of the cross section in Eq.\ (39) suggests that in a
    certain gauge or in some gauge invariant way it could be written
    in a factorized form involving two unintegrated gluon
    distributions merged by an effective Lipatov vertex}.
\end{quote}
There is no derivation of Eq.\ \eqref{ktfact}.  Rather,
this equation is stated as being the ``usual form  
of the factorized inclusive cross section'', with the source being our
Refs.\
\cite{Kovner:1995ja, Kovner:1995ts, 
Kovchegov:1997ke, Guo:1998pe, Gyulassy:1997vt}. This is indicated by
the arrows away from Ref.\ \cite{Kovchegov:2001sc} in Fig.\
\ref{citationgraph}.   
Furthermore, the paper then goes on to say that the results of the
calculations actually appear to be ``in \emph{disagreement} with the
factorization \emph{hypothesis}''.  (Our emphasis.)  That is, Eq.\
\eqref{ktfact} is actually false in the situation considered in Ref.\
\cite{Kovchegov:2001sc}. Moreover, the calculations in the paper are
for gluon production in DIS, whereas the papers that cite Ref.\
\cite{Kovchegov:2001sc} as a source for the factorization formula
employ it in hadron-hadron collisions.  Thus Ref.\
\cite{Kovchegov:2001sc} seems to be a particularly bad reference to
cite for the validity of $k_T$-factorization, especially in
nucleus-nucleus scattering, as is the case in \cite{Kharzeev:2004if,
  Levin:2010dw, Levin:2010zy, Albacete:2010bs, Albacete:2010ad,
  Rezaeian:2011ia}.

To be fair, in a later paper \cite{Kharzeev:2003wz}, the authors of
\cite{Kovchegov:2001sc} do propose a solution to the problem of the
inconsistency of their calculations with $k_T$-factorization.  The
solution is that a different definition of the gluon distribution is
needed --- see the section of \cite{Kharzeev:2003wz} that is titled ``A
tale of two gluon distribution functions''.  But this simply
reinforces our statement
that \cite{Kovchegov:2001sc} is an entirely inappropriate reference to
cite for $k_T$-factorization and its validity, at least not without
considerable further explanation.  

We next examine the sources \cite{Kovner:1995ja, Kovner:1995ts,
Kovchegov:1997ke, Guo:1998pe, Gyulassy:1997vt} 
that Ref.\ \cite{Kovchegov:2001sc} itself
cites for the factorization formula Eq.\ \eqref{ktfact}:
\begin{itemize}

\item In Refs.\ \cite{Kovner:1995ja,Kovner:1995ts} there actually exists no formula at
all resembling \eqref{ktfact},  
and those papers are not concerned with proving factorization. We indicate this fact in Fig.\ \ref{citationgraph}
by placing a cross on the references to
\cite{Kovner:1995ja,Kovner:1995ts}. 

\item
In Ref.\ \cite{Kovchegov:1997ke}, there is one equation, (40), that has
a somewhat similar structure to our \eqref{ktfact}.  But in detail it
is quite different, for example as regards the overall normalization
and the number of powers of the strong coupling $g$, and the meaning
of the functions inside the transverse-momentum integral.  Thus this
reference cannot be used to support a statement that \eqref{ktfact} is
the ``usual form of the factorized inclusive cross section''.

\item 
Reference \cite{Guo:1998pe} is concerned with a possible interpretation of the classical formula 
within perturbation theory using the DGLAP splitting kernels, but Eq.\ \eqref{ktfact} is never 
present in the reference and we therefore again indicate this fact by placing a cross 
on the reference in Fig.\ \ref{citationgraph}.  

\item
Finally, Ref.\ \cite{Gyulassy:1997vt} does indeed contain
Eq.\ \eqref{ktfact}, which is referred to as the ``GLR formula'' and is given by 
Eq.\ (65) in the reference. 
The origin of the formula is given as a review article by Gribov, Levin and Ryskin \cite{Gribov:1984tu} (GLR),
as indicated in Fig.\ \ref{citationgraph}.  

\end{itemize}
Thus to trace the origin of the $k_T$-factorization formula \eqref{ktfact}
we have come from the recent applications down to the GLR review
\cite{Gribov:1984tu}, via Ref.\ \cite{Kovchegov:2001sc}. 

So now let us turn to Ref.\ 
\cite{Gribov:1984tu}.  There the $k_T$-factorization
formula appears as Eq.\ (4.3).  But the numerical coefficient is
$N_c/(2\pi)^6$, which is extremely different from the $2/C_F$ in
our \eqref{ktfact} and in Eq.\ (40) in \cite{Kovchegov:2001sc}.
In fact, as discussed in Sect.\ 4.4.1 of \cite{Avsar:2012hj}, there 
appear many different formulas for the coefficient of \eqref{ktfact}. The choice in \eqref{ktfact} 
corresponds to the choice in \cite{Kovchegov:2001sc} (and the majority of papers use this choice, as explained in 
section 4.4.1 of \cite{Avsar:2012hj}).

Indeed the normalization issue is referred to in the paper
\cite{Gyulassy:1997vt} that is citationally nearest to the GLR paper,
in a section entitled ``Comparison with GLR formula''.  But the
combination of (65) and (68) in \cite{Gyulassy:1997vt} agrees with our
\eqref{ktfact} except for a previously mentioned redefinition of gluon
densities by a factor of $k_T^2$.  No mention is made of the
dramatically different normalization factor in the GLR paper, even
though the GLR paper is quoted as the source for the $k_T$-factorization
formula.  (Note that the formula in \cite{Gyulassy:1997vt}, like our
\eqref{ktfact}, is for $d\sigma/d^2p_T dy$, whereas the formula in GLR
is for the apparently different quantity $Ed\sigma/d^3p$; however,
both these quantities are equal after applying the appropriate change
of kinematic variable.)

In addition to the issue of normalization there are several
other relatively minor differences between the versions of
formula \eqref{ktfact} among different papers. In
\cite{Kharzeev:2004if} for example, the $k_\perp$ integral has $p_\perp$ as
the upper limit, while other papers do not indicate such a limit.  The
use of such an upper limit is rather non-trivial because it does not
treat the hadrons in a completely symmetric manner.  Moreover, the
integration measure is sometimes written $d^2k_\perp$ (e.g.
\cite{Kovchegov:2001sc, Kharzeev:2003wz, Armesto:2004ud}) and
sometimes $dk_\perp^2$ (e.g. \cite{Gribov:1984tu, Kharzeev:2004if}) so it
is not clear what the correct prescription is. Finally, the coupling
$\alpha_s$ is treated differently from paper to paper. In
\cite{Gribov:1984tu} it is put inside the integral, with a
$k_T$-dependent scale, while other papers write it as in
\eqref{ktfact}, with a fixed $\alpha_s$ outside the integral.  For someone
well experienced with this area, these differences are likely to be
unimportant.  Thus the difference in the arguments of $\alpha_s$ could signal
the difference between a strict leading-logarithm approximation and an
idea about the likely result of an improved approximation.

But for an outsider to the topic, these differences can be quite
bothersome, because it is not clear how they arise, or how
significant they are, particularly when there is no comment on the
differences.  This is especially true in the absence of any
derivations where one can locate the source of the differences.

At this point, an outsider should feel entitled to be able to examine
a detailed step-by-step derivation, so as to be able to pinpoint where
the error(s) is/are.  That there is an error somewhere is absolutely
demonstrated by the difference in normalization.  As we have explained,
there are no relevant derivations in the chronologically later papers
whose citation 
chains led us back to Ref.\ \cite{Gribov:1984tu}.  But in \cite{Gribov:1984tu},
we find no detailed derivation either. Just above that paper's (4.3) there is
a statement of what graphs are involved (with only a very brief
explanation) and then a statement of what is supposed to be the
resulting factorization formula.  However, one is referred back to
two previous papers \cite{Gribov:1981kg, Gribov:1983fc} by the same
authors, as indicated in Fig.\ \ref{citationgraph}.  

But in these papers, we merely find that there is an assertion that
certain graphs dominate, an assertion with a stated reason that
certain other graphs cancel, and then a statement of the resulting
formula.  There is no detailed derivation, e.g., a
derivation that is sufficiently detailed to make it manifest, for
example, how the many factors of $2\pi$ that are ubiquitous in loop
integrals organize themselves to give one or other of the
very different normalization factors that are found in different
version of the $k_T$-factorization formula.

It should also be mentioned that the assertions made in Refs.\ \cite{Gribov:1983fc, Gribov:1984tu} are based 
on the axial gauge where it is
known that the unphysical singularities introduced in the gauge propagators 
do cause problems with
divergences in the most natural simple definitions of the TMD parton  
densities, and additionally the contour deformations needed to prove 
factorization are blocked by these extra poles. See, e.g., \cite[Chs.\
13 \& 14]{qcdbook} for details of this.
We find no indication in the 
cited references that such technical (yet crucial) issues are addressed. 

In \cite{Gribov:1983fc}, we read that the equation (with the
normalization given in GLR, not the normalization that is currently
accepted) ``can easily be obtained in the QCD leading logarithmic
approximation'', and the relevant references given are
\cite{Gribov:1981kg, Ryskin:1980yz}.

So we are now lead back to the papers \cite{Gribov:1981kg,
Ryskin:1980yz}. These two references are indicated by the red boxes in
Fig.\ \ref{citationgraph} and they thus constitute the two sources
where one finally might expect to find a proof of \eqref{ktfact}, but
presumably with a different normalization factor.  

We first examine \cite{Gribov:1981kg}, in which we have been told in
\cite{Gribov:1983fc} that a derivation is to be found.  Again we find
no derivation.  There is again just a statement of the diagrams
involved and a statement of the resulting factorization formula.
There is no derivation that can be debugged to check the
normalization; there is not even any justification of why the
particular graphs dominate.  Neither is any statement given of the
approximations involved.  No prior references are given.  
Similarly in \cite{Ryskin:1980yz} we again find a short statement  of the diagrams
involved and of the factorization formula, but no detailed
derivation or justification. 

In \cite{Gribov:1981kg} it is also mentioned, below that paper's Eq.\ (1), 
that the authors managed to calculate the relevant diagrams for the high-$p_T$ 
production of hadrons in Ref.\ \cite{Gribov:1980uj}. We therefore also include this reference 
in Fig.\ \ref{citationgraph}. 
In Ref.\ \cite{Gribov:1980uj}, the $k_T$-factorization formula is presented
in the very last equation of the paper (un-numbered). It is, however,
again simply stated without any derivation. Nor is any further
reference or argument provided, and it is not at all clear what the
definitions of the parton distributions are. It should also be noted
that in \cite{Gribov:1980uj}, an axial gauge is used for studying
the leading graphs of high-$p_T$ processes.  Thus one would here again
expect the problems mentioned above that stem from unphysical gauge
singularities.  We find again no indication in the references that these
problems are dealt with.

In Fig.\ \ref{citationgraph}, we have enclosed the boxes
\cite{Gribov:1980uj, Ryskin:1980yz, Gribov:1981kg} in thick red lines.  This denotes
that they are at the end of the chains of citations, and that the
derivations we should expect to find in them do not exist.

In the appendices of the present paper, we present a derivation of a
$k_T$-factorization formula starting from what appear to be the same
assumptions and approximations used by GLR.  We get a quite different
formula, as regards the normalization.  Our proof gives what we intend
to be enough detail that an outsider can verify the result.  The
assumptions and approximations are valid in a model that uses
lowest-order graphs, so that our derivation is adequate to test
whether a claimed result is actually true.

A summary of this section is that we find that regarding the important and widely used factorization formula \eqref{ktfact},
the principles (T1)--(T4) are violated very badly. A confusing chain of citations is present in the literature 
as shown in Fig.\ \ref{citationgraph}.  By following the relevant references we are unable to find any paper 
where a proper derivation is provided. 
Moreover, the explicit definitions of the basic entities involved are not always clear, 
and differ from paper to paper (see also section 4 of \cite{Avsar:2012hj} for a more explicit discussion).  
It is therefore extremely important for progress in the field that all of recommendations (T1)--(T4) are adapted and maintained universally.

\section{Summary}
\label{sec:summary}

Our principles (T1)--(T4) embody sensible standards and conventional
norms for the presentation of work in theoretical physics.  Suppose a
paper uses, but does not derive, a particular formula of the status of
the $k_T$-factorization formula \eqref{ktfact} as a key part of its
work.  Then it is a normal and reasonable expectation that a reference
should be given to where the formula is derived or otherwise
justified.  This we found to be badly violated, as summarized in Fig.\
\ref{citationgraph}.

For example, a main reference \cite{Kovchegov:2001sc} frequently given
for \eqref{ktfact} not merely failed to derive it but actually
questioned its validity in the relevant context.  Moreover, of the
references that \cite{Kovchegov:2001sc} gives for the formula, most
failed to even contain the formula.  Backtracking through a chain of
references where the formula does appear we eventually met a dead end:
we could not actually find a useful derivation.  

This is a quite disturbing situation.  The formula in question is a
key part of an important area of QCD phenomenology, and a superficial
reading of the literature indicates that the formula appears to be
accepted without question as the standard one.  But closer
examination fails to turn up a justification, at least not by
following the normal and uncontroversial procedure of examining the
references cited for the formula.

Of course, it may be that by a much wider search through the
literature one can find an adequate justification of \eqref{ktfact}.
But the point of a citation is to avoid a broad literature search: The
reader just needs to go to the cited article(s) and find both the
formula and its justification.  Any substantial deviations from this
expectation need explicit explanation.

A failure to find appropriate justification of a mainstream formula
immediately raises the question as to whether the formula is correct.
Recent work \cite{Collins:2007nk, Rogers:2010dm, Forshaw:2012bi} finds violations of
standard factorization in related situations, so ignoring the problems
is both dangerous and can lead to failure to find correct consequences
of QCD.

For a single paper whose citations are inadequate, it may be
most appropriate to contact the author(s) to get corrected
information.  But in the present case, the problems are much more
widespread, and need a more widespread concerted remedy. We also 
emphasize that we have identified other similar problems in the justification 
of related topics but for the sake  of brevity we document here only the situation for 
one case: $k_T$-factorization for gluon production in hadron-hadron collisions. 

We conjecture that one reason for the inadequate citations is that the
$k_T$-factorization formula is regarded by its users as
uncontroversial.  Thus providing citations for it is a formality
rather than a serious scientific exercise.\footnote{%
   \label{fn:mismatch1}%
   In addition, the
   foundational work on a topic, which may be decades old, may give the
   key ideas, but not the extra steps to put results in the form in which
   they are later found to be most useful and convenient.  It may be
   that these extra steps are known privately to insiders but that
   they do  not appear in the papers 
   that it is most natural to cite.  We give an explicit example in Sec.\
   \ref{sec:gluon.density.def}. This phenomenon can explain the problems
   with the citations, but it leaves unchanged the readers' difficulty of
   reproducing results.}  We conjecture that a second reason is that
there is substantial pressure, particularly on young scientists, to
produce new results, and that it is regarded as counterproductive to
actually check the validity of earlier results that are relied on to
obtain the new results.

In addition, a disincentive to giving a full derivation is its length,
if one makes explicit the details needed by a newcomer.  For example,
the derivation in the appendix of the present paper adds about $50\%$
to the paper's length.

Our findings show that such tendencies need to be reduced for the
health of our field; they greatly and unnecessarily impede the ability
of other physicists to reproduce and verify the work.  Insiders may
have good reason to trust the formula, but workers in even closely
related areas of QCD cannot check the results, at least not at all
easily. The responsibility is not only on authors to justify their
results adequately, but also on journal referees to check that results
quoted in a new paper have actually been justified.  Note that the
problems we are finding are not at the level of a deep critique of a
difficult proof; we find there is such an absence of derivation that
there is not even a proof to critique.

\section*{Acknowledgments}

This work was supported in part by the U.S. Department of Energy under
grant number DE-FG02-90ER-40577.  We thank Markus Diehl, Ted Rogers,
and Mark Strikman for very useful comments on a draft of this paper;
the current version includes responses to some of their questions.

\appendix
\section{Derivation of the normalization factor of the
  $k_T$-factorization  formula}

Since we have been unable to find a derivation of the
$k_T$-factorization formula \eqref{ktfact} by following references for
it, and since there is disagreement on the normalization factor, we
provide here a simple derivation that gives the normalization.  

The derivation starts from certain assumptions concerning the graphs
and momentum regions that dominate.  Since the assumptions are
demonstrably incorrect at sufficiently high order in perturbation
theory, as we will explain, the derivation cannot be completely
correct as regards full QCD.  Nevertheless, the derivation does apply
to a simple Feynman graph model, and therefore unambiguously
determines the appropriate value for the normalization, which differs
from both of those in the standard references, \cite[Eq.\
(4.3)]{Gribov:1984tu} and \cite[Eq.\ (40)]{Kovchegov:2001sc}.
Moreover, the assumptions we make are compatible with the stated
starting point given in \cite{Gribov:1984tu, Gribov:1981kg,
  Gribov:1983fc}, so our derivation can be used to test the
correctness of the factorization formula stated in those references.

Our derivation is a simplified version of a derivation that one of us
gave in Ref.\ \cite{Avsar:2012hj}.

\subsection{Definition of process and gauge}

We consider the cross section $d\sigma/d^2p_Tdy$ for inclusive
production of a gluon in a high-energy collision of unpolarized
hadrons, where the gluon has transverse momentum $p_T$ and rapidity
$y$. The kinematic region of interest is where $\Lambda \ll |p_T| \ll
\sqrt{s}$ and $e^{y_B} \ll e^y \ll e^{y_A}$, where $\sqrt{s}$ is the
center-of-mass energy, and $y_A$ and $y_B$ are the rapidities of the
beam hadrons. We define $Q=|p_T|$, and let the momenta of the
incoming hadrons be $P_A$ and $P_B$.

The aim is to obtain a factorized form for the cross section that is
valid up to corrections that are a power of the small parameters
$p_T/\sqrt{s}$, $e^{-(y-y_B)}$, and $e^{-(y_A-y)}$. 

We use light-front coordinates relative to the two beams, with the
momentum of the detected gluon being
\begin{equation}
  p = \left( p^+, p^-, p_T \right)
    = \left( \frac{Qe^y}{\sqrt{2}}, \frac{Qe^{-y}}{\sqrt{2}}, p_T \right) .
\end{equation}

Following \cite{Collins:1981uw,Collins:1981uk}, we use a time-like
axial gauge $n\cdot \phi=0$.  Here the symbol $\phi$ denotes the gluon field,
since we use the usual symbol $A$ for another purpose.
The gauge-fixing vector has the
same rapidity as the detected gluon:
\begin{equation}
  n \propto (e^y, e^{-y}, 0_T),
\end{equation}
Then the numerator of the gluon propagator has the form
\begin{equation}
\label{eq:axial.gauge.num}
  -g^{\mu\nu} + \frac{ k^\mu n^\nu + n^\mu k^\nu }{ k \cdot n}
  - \frac{ k^\mu k^\nu n^2 }{ (k \cdot n)^2 }.
\end{equation}
The point of this gauge is that for gluons of high positive rapidity
it approximates the light-like axial gauge $\phi^+=0$ while for gluons of
high negative rapidity it approximates the light-like axial gauge
$\phi^-=0$.  This happens since $k \cdot n \simeq k^+n^-$ when the
rapidity of $k$ is large and positive, while $k \cdot n \simeq k^-n^+$
when the rapidity of $k$ is large and negative.

When various complications are ignored, each of these different
light-like gauges gives an appropriately simple formulation of
parton-model physics inside the corresponding beam hadron --- see
\cite[Sec.\ 7.4]{qcdbook} and references therein.

\subsection{Assumptions}

\begin{figure}
  \centering
  \includegraphics[scale=0.5]{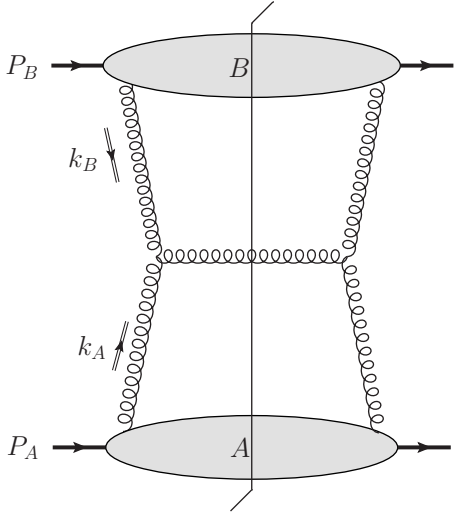}
  \caption{Graphs assumed to dominate for Eq.\ \protect\eqref{ktfact}.}
  \label{fig:dominate}
\end{figure}

We make the following assumptions:
\begin{enumerate}
\item The cross section is dominated by (cut) graphs of the form of
  Fig.\ \ref{fig:dominate}, where one gluon out of each hadron
  combines to form the observed gluon.  Lines associated with the
  incoming hadrons are in the subgraphs $A$ and $B$.
\item We assume that the rapidities of lines in $A$ are substantially
  greater than the rapidity $y$ of the observed gluon, i.e., the
  rapidities in $A$ are at least $y+\Delta$, with $e^\Delta \gg 1$.
  This and the next assumption encode a key kinematic property used in
  BFKL physics.
\item Similarly the rapidities in $B$ are less than
  $y-\Delta$. 
\item Transverse momenta in $A$ are of order $Q$ or smaller, and
  transverse momenta in $B$ are of order $Q$ or smaller.
\item We ignore the fact that a massless gluon is not a physical
  particle in real QCD, as opposed to low-order perturbation theory.
\item We ignore all other subtleties, e.g., that certain other
  graphical structures may be important, and can even break
  factorization \cite{Collins:2007nk, Rogers:2010dm, Forshaw:2012bi}.
\end{enumerate}

\begin{figure}
  \centering
  \includegraphics[scale=0.5]{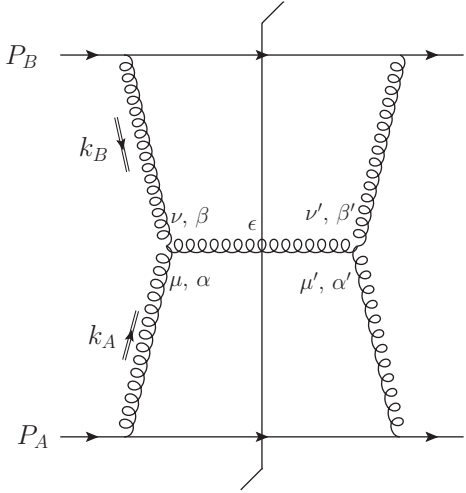}
  \caption{Lowest order example of Fig.\
    \protect\ref{fig:dominate}. Here $\mu$, $\alpha$, etc, are Lorentz
    and color indices for the gluons attaching to the horizontal gluon
    line. }
  \label{fig:LO}
\end{figure}

The assumptions are valid for the lowest-order graph in a perturbative
model of high-energy quark-quark scattering, Fig.\ \ref{fig:LO}.  In
this graph, it is readily checked that the kinematic assumptions for
the $A$ and $B$ subgraphs follow from the kinematic region chosen for
the detected gluon.  It can also be checked that other graphs \emph{of
  this order} are power-suppressed in the gauge we have chosen, given
that polarization vectors $\epsilon$ for the final-state gluon obey
$\epsilon \cdot p = \epsilon \cdot n = 0$.

\subsection{Cross section}

At high-energy, the cross section from Fig.\ \ref{fig:dominate} is
\begin{multline}
  \label{eq:xsect.basic}
  \frac{ d\sigma }{ d^2p_Tdy }
  = \frac{1}{2s}
    \int \frac{ d^4k_A \, d^4k_B }{(2\pi)^4} \, \delta^{(4)}(k_A+k_B-p) 
\times \\ \times
   \frac{1}{16\pi^3}
    A^{\mu\mu',\alpha\alpha'} B^{\nu\nu', \beta\beta'}
    H_{\mu\mu',\nu\nu'; \alpha\alpha', \beta\beta'}.
\end{multline}
Here, $A^{\mu\mu',\alpha\alpha'}$ is the amplitude for the subgraph
$A$ including the propagators for its two external gluons, as a
function of the Lorentz and color indices of these gluons.  It
includes integrations over internal loop momenta and the momenta of the
final-state particles of $A$.  
Similarly, $B^{\nu\nu', \beta\beta'}$ is for subgraph $B$.
The factor
\begin{equation}
  \frac{1}{16\pi^3}
\end{equation}
arises from the Lorentz-invariant differential for the detected gluon:
\begin{equation}
\label{eq:LIPS}
  \frac{d^3p}{(2\pi)^32E_p}
  = \frac{dy d^2p_T}{16\pi^3},
\end{equation}
given that the cross section is differential in $p_T$ and $y$.
Finally, $H_{\mu\mu',\nu\nu'; \alpha\alpha', \beta\beta'}$ is the
factor for the central gluon:
\begin{multline}
\label{eq:H}
  H_{\mu\mu',\nu\nu'; \alpha\alpha', \beta\beta'}
  = 
  \sum_\epsilon g^2 f_{\alpha\beta\gamma}f_{\alpha'\beta'\gamma}
\times \\ \times 
  \left[ (k_A-k_B)\cdot\epsilon g_{\mu\nu}
         + (k_B+p)_\mu \epsilon_\nu
         + (-p-k_A)_\nu \epsilon_\mu
   \right]
\times \\ \times 
  \left[ (k_A-k_B)\cdot\epsilon g_{\mu'\nu'}
         + (k_B+p)_{\mu'} \epsilon_{\nu'}
         + (-p-k_A)_{\nu'} \epsilon_{\mu'}
   \right].
\end{multline}
The sum is over two independent orthonormal polarization vectors for
the gluon.  In a frame in which $p_T$ is along the $x$-axis:
\begin{equation}
  p = Q \left( \frac{e^y}{\sqrt{2}}, \frac{e^{-y}}{\sqrt{2}}, 1, 0 \right) ,
\end{equation}
we can choose
\begin{equation}
  \epsilon_1 = (0,0,0,1)
  ~\mbox{and}~
  \epsilon_2 = \left( e^y, -e^{-y}, 0, 0 \right)/ \sqrt{2}
\end{equation}
as the independent polarization vectors.

To keep track of the dependence of calculations on the gauge group, we
will assume that the gauge group is SU($N_c$).

\subsection{Leading-power approximations}

We now make approximations suitable for the kinematic region that we
have assumed.  

First we need the size of the components of the momenta
$k_A$ and $k_B$ of the gluons coming out of $A$ and $B$:
\begin{align}
\label{eq:kA.size}
  k_A &= \left( \frac{Qe^y}{\sqrt{2}} + O(Qe^{y-\Delta}), ~
                O(Qe^{-y-\Delta}), ~
                k_{A,T}
         \right),
\\
\label{eq:kB.size}
  k_B &= \left( O(Qe^{y-\Delta}), ~
                \frac{Qe^{-y}}{\sqrt{2}} + O(Qe^{-y-\Delta}), ~
                k_{B,T}
         \right).
\end{align}
The components $k_A^-$ and $k_B^+$ are small relative to the
corresponding components of $p$, since the components of $k_A$ and
$k_B$ cannot exceed the typical values for momenta in $A$ and $B$;
the sizes of momenta in $A$ and $B$ are governed by our assumed
separation of rapidity between $A$, $B$ and the gluon $p$.  Then
momentum conservation at the three-gluon vertex ensures that $k_A^+$
and $k_B^-$ equal $p^+$ and $p^-$, up to small corrections.

Next consider the line for one of the external gluon lines of $A$. It
connects a subgraph $H$ characterized by rapidity $y$ to a subgraph
$A$ with rapidity $y+\Delta$, and we write the relevant part of the
graph as
\begin{equation}
\label{eq:AH}
  a^\lambda \, N_{\lambda\mu}(k_A) \, h^\mu.
\end{equation}
Here $N_{\lambda\mu}$ is the numerator of the gluon propagator, while
$a$ and $h$ correspond to the rest of the $A$ subgraph and to the $H$
subgraph. We treat $h^\mu$ and $a^\lambda$ as being obtained by a boost
from a situation with zero rapidity, and therefore we obtain relative
sizes for the components:
\begin{align}
  a =  (a^+,a^-,a_T) & \sim  \mbox{const.} \times (e^{y+\Delta}, e^{-y-\Delta}, 1),
\\
  h & \sim \mbox{const.} \times (e^y, e^{-y}, 1). 
\end{align}
We use the sizes given in \eqref{eq:kA.size} to determine that the sum
over components in \eqref{eq:AH} is dominated by $\lambda=+$ and the
transverse components in $\mu$:
\begin{equation}
\label{eq:dominant.pol}
  - a^+ N^{-i} h^i.
\end{equation}
A corresponding calculation is then applied to each of the gluons
connecting the central gluon to the $B$ subgraph.  

Finally we observe that $k_A^+$ and $k_B^-$ differ from $Qe^y/\sqrt2$
and $Qe^{-y}/\sqrt2$ by an amount that is small compared with the
corresponding components of momenta in $A$ and $B$.  Thus in $A$ and
$B$ we can replace $k_A^+$ by $Qe^y/\sqrt2$, and $k_B^-$ 
by $Qe^{-y}/\sqrt2$.

We then find that, given our assumptions, the cross section is given
by
\begin{multline}
  \label{eq:xsect.approx1}
  \frac{ d\sigma }{ d^2p_Tdy }
  = \frac{\pi}{2s}
    \int d^2k_{A,T}
    \int \frac{dk_A^-}{(2\pi)^4}
    \int \frac{dk_B^+}{(2\pi)^4}
\times \\ \times 
    A^{ii',\alpha\alpha'}
    B^{jj', \beta\beta'}
    H^{ii',jj'; \alpha\alpha', \beta\beta'},
\end{multline}
where the integrals and factors of $2\pi$ are arranged to correspond
to those in the definition of a gluon density, as we will see.

\subsection{Gluon density}
\label{sec:gluon.density.def}

If there were no complications, the natural method to define a number
density of a parton is by the hadron expectation value of the number
operator in the sense of light-front quantization.  The basic ideas
were worked out by Bouchiat, Fayet, and Meyer \cite{Bouchiat:1971mj}
and by Soper \cite{Soper:1977jc}, although these papers do not
contain actual formulas for transverse-momentum-dependent
parton densities in terms of hadron matrix elements of field
operators.\footnote{This gives an example of the difficulty referred
  to in footnote \ref{fn:mismatch1}}  In light-front quantization and
light-front gauge, this gives the gluon density as
\begin{multline}
  P(x,k_T) 
  =
  \frac{xP^+}{(2\pi)^3}
  \int dw^- d^2w_T \, e^{-ixP^+w^- + i k_T\cdot w_T}
\times \\ \times 
  \langle P| \phi^{i,\alpha}(0,w^-,w_T) \phi^{i,\alpha}(0) | P \rangle,
\end{multline}
where $\phi^{\mu,\alpha}$ denotes the gluon field operator.
This definition is equivalent to Eq.\ (2.3) of \cite{Collins:1981uw} when the
light-cone gauge $\phi^+=0$ is used, which we will now show to be
appropriate in the approximation we are using.

If gluon momenta all have high
rapidity, then $k\cdot n = k^+n^- \left( 1 + \frac{k^-}{k^+}
  \frac{n^+}{n^-} \right) \simeq k^+n^-$ in the gluon propagator, where the
neglect of $k^-n^+/(k^+n^-)$ relative to unity follows because the
rapidity of $k$ is much higher than that of $n$.  Note that for the
gluon momentum $k_A$ entering the hard scattering, the size of the
plus component is appropriate 
for the hard scattering, i.e., $k_A^+ \sim Qe^y$, but the minus component
is appropriate for subgraph $A$, i.e., $k_A^- \sim Qe^{-y+\Delta}$, so that
$k_A$ has effectively rapidity of about $y+\Delta$. 

Hence in the gluon
density, time-like axial gauge is equivalent to $\phi^+=0$ gauge to
leading power. The above definition then gives
\begin{equation}
  P(x,k_T) 
  =
  k^+ \int \frac{ dk^- }{ (2\pi)^4 }
  A^{ii,\alpha\alpha}(k,P).
\end{equation}

The normalization of the above definition is that
$P(x,k_T)$ would be a number density of gluons in $x$ and $k_T$, if we
ignored the complications present in higher order corrections in
QCD. The corresponding quantity in the $k_T$-factorization formula
\eqref{ktfact} is actually a momentum density.  Therefore, in the
above formulas, we follow Refs.\ \cite{Soper:1979fq, Collins:1981uw},
and use $P(x,k_T)$ to denote the number density, rather than the
more common notation $f(x,k_T)$.

Of course, as explained in \cite[Chs.\ 13 \& 14]{qcdbook} the
light-cone-gauge definition\footnote{Note that the definition in
  \cite{qcdbook} is for a TMD quark density.  For a TMD gluon density,
  one makes the same adjustments as are used to get the formula for an
  integrated gluon density from the formula for an integrated quark
  density.} has problems, because of rapidity divergences and other
issues, and even the definition with a non-light-like axial gauge has
trouble.  So the definitions have to be modified to be suitable for
full QCD.  But within the approximation used here, the simplest
definition is adequate.

\subsection{Factorization}

We next recall that in the small-$x$ limit and in the approximation
we use here, only the term \eqref{eq:dominant.pol} in the lines for
the external gluons of $A$ is important.  So for each gluon line there
is a factor of the vector transverse momentum multiplying a
rotationally invariant quantity.  Moreover with a color-singlet or a
color-averaged beam, the dependence of $A$ on color is proportional to
$\delta_{\alpha\alpha'}$.

Thus for the $A$ factor in \eqref{eq:xsect.approx1}, we have
\begin{align}
    \int \frac{dk_A^-}{(2\pi)^4}
    A^{ii',\alpha\alpha'}
& =
  \frac{ k_A^ik_A^{i'} }{ |k_{A,T}|^2 }
  \frac{ \delta_{\alpha\alpha'} }{ N_c^2-1 }
    \int \frac{dk_A^-}{(2\pi)^4}
    A^{jj,\beta\beta}
\nonumber\\
& = \frac{ k_A^ik_A^{i'} }{ |k_{A,T}|^2 }
  \frac{ \delta_{\alpha\alpha'} }{ N_c^2-1 }
  \frac{1}{ k_A^+ } P(x_A, k_{A,T} ).
\end{align}
Here the factors of $k_A^i$ and $k_A^{i'}$ arise from the axial-gauge
gluon propagator as it is used in \eqref{eq:dominant.pol}, while
$x_A=Qe^y/(\sqrt{2} p_A^+) = Qe^y/\sqrt{s}$ (in the overall
center-of-mass frame).  An exactly similar result applies to the $B$
subgraph, with $x_B=Qe^{-y}/(\sqrt{2} p_B^-) = Qe^{-y}/\sqrt{s}$.
Substituting these results in \eqref{eq:xsect.approx1} gives
\begin{multline}
  \frac{d\sigma}{d^2p_T dy} 
  = \frac{\pi}{x_Ax_Bs^2 (N_c^2-1)^2 } 
    \int d^2k_{A,T} \, P_A(x_A,k_{A,T}) 
\times \\ \times 
    P_B(x_B, p_T - k_{A,T})
     \frac{ k_A^ik_A^{i'} k_B^jk_B^{j'} H^{ii',jj'; \alpha\alpha, \beta\beta} }
          { |k_{A,T}|^2 |k_{B,T}|^2 }.
\end{multline}
Performing the algebra in the $H$ factor, and making approximations
valid to leading power in $e^{-\Delta}$ then gives the following
factorization formula
\begin{multline}
  \label{eq:ktfact.correct}
  \frac{d\sigma}{d^2p_T dy} 
  = \frac{2\pi^2\alpha_s}{C_F \, p_T^2} 
    \int d^2k_{A,T} \, x_AP_A(x_A,k_{A,T}) 
\times \\ \times
    x_BP_B(x_B, p_T - k_{A,T}),
\end{multline}
with $\alpha_s=g^2/(4\pi)$ and $C_F=(N_c^2-1)/(2N_c)$, as usual.
Notice that each gluon density has been multiplied by a factor of the
corresponding $x$ value, and that this compensated a change from $1/s$
to $1/p_T^2$ in the pre-factor.

This factorization formula is to be compared with the formula stated
earlier, \eqref{ktfact}, which has the normalization given in
\cite[Eq.\ (40)]{Kovchegov:2001sc}.  Evidently in that formula, the
gluon densities are to be interpreted as momentum densities rather
than number densities, i.e., $f(x,k_T)=xP(x,k_T)$.  But there is also
a factor of $\pi^2$ missing.

\subsection{Why the proof is insufficient}

Our derivation of the factorization formula \eqref{eq:ktfact.correct}
relied on certain assumptions, which are not in general true in
QCD. However they are true for the lowest-order graph, so that our
derivation gives the correct normalization factor.  At a minimum the
proof needs enhancement if the result is to be regarded as actually
derivable in full QCD.  We now explain some of the difficulties.

In the first place the definition of the gluon density needs to be
changed.  Already, it was seen in \cite{Collins:1981uw,Collins:1981uk}
that the natural definition of a TMD quark density in light-cone gauge
has rapidity divergences.  This was corrected by the use of non-light
axial gauge.  But even this has problems \cite[App.\
A]{Bacchetta:2008xw}.  A satisfactory gauge-invariant definition for a
TMD quark density was recently given by one of us in \cite[Chs.\ 13 \&
14]{qcdbook}, and the definition can be extended in a reasonably
natural way to a TMD gluon density.  However, the treatment in
\cite{qcdbook} did not cover the additional issues that arise at small
$x$.

In general, many more kinds of graph and momentum region contribute.
If factorization is to hold, various summations and cancellations have
to be employed.  Moreover some of the cancellations involve issues of
relativistic causality --- see \cite{Bodwin:1985hc,Collins:1988ig} ---
which is broken graph-by-graph by the $k\cdot n$ denominators in the
axial-gauge gluon propagator --- see \eqref{eq:axial.gauge.num}.  Use
of the Feynman gauge increases the number of graphical structures and
regions to be considered.  

Furthermore, it has recently been found that factorization is actually
violated in some significant situations in hadron-hadron collisions
\cite{Collins:2007nk, Rogers:2010dm, Forshaw:2012bi}.  While these
particular examples do not include the particular process discussed in
the present paper, they do show that claimed proofs of factorization
need careful attention to ensure that they actually work.

\section{Testing the normalization factor}

The argument that derives the normalization of the $k_T$-factorization
formula is somewhat long.  Since the result disagrees with both of the
standard formulas to be found in the literature, it is useful to find
some elementary tests to verify our result.

In our formula \eqref{eq:ktfact.correct} the factors of $\pi^2$ are
different from those in both \cite[Eq.\ (4.3)]{Gribov:1984tu} and
\cite[Eq.\ (40)]{Kovchegov:2001sc}.  The group theory factor $1/C_F$
does agree with that in \cite{Kovchegov:2001sc}, but disagrees with the
factor $N_c$ in \cite{Gribov:1984tu}, which is written there simply as
$N$. 

We do the verification by determining how the powers of $\pi$ and the
$N_c$ dependence arise in the lowest-order graph Fig.\ \ref{fig:LO}
for the cross section and in the lowest-order graph for the gluon
density.  We will show how this immediately gives the power of $\pi$
and the $N_c$ dependence in the prefactor in the factorization formula
\eqref{eq:ktfact.correct}.  In the lowest-order graphs and in $n\cdot
\phi=0$ gauge, all our assumptions are valid, and agree with the starting
point in GLR \cite{Gribov:1984tu}, so at lowest-order there is no
ambiguity as to the correct result.

\subsection{Power of $\pi$}

In the cross section, the integral over the plus and minus components
of the loop momentum are performed using the mass-shell
delta-functions for the spectator quarks.  After the standard
kinematic approximations these give simple algebraic calculations,
and hence no factors of $\pi$.  There remains a two-dimensional
transverse momentum integral which is unchanged between the cross
section calculation and the factorization formula.  From the
differential \eqref{eq:LIPS}, there is a $1/\pi^3$ for each of the 3
final-state particles, and there is a $\pi^4$ associated with the
delta-function for overall momentum conservation.  This gives an
overall factor of $1/\pi^5$ in the cross section.

In the one-loop gluon density the definition itself contains an
overall factor of $1/\pi^4$, and there is a factor of $\pi$ from the
$2\pi\delta((p-k)^2)$ that puts the spectator quark on-shell.  This
gives an overall factor of $1/\pi^3$.

To match these powers of $\pi$, the coefficient of the two gluon
densities in the factorization formula for the cross section has a
factor of $\pi$.  Since $\alpha_s = g^2/(4\pi)$, we need a factor
$\pi^2$ multiplying $\alpha_s$, which is exactly what we have in
\eqref{eq:ktfact.correct}.  The factorization formulas in both 
of Refs.\ \cite{Gribov:1984tu, Kovchegov:2001sc} have a different
power of $\pi$ and are therefore wrong.

\subsection{Dependence on $N_c$}

We now ask for the dependence on $N_c$ when the gauge-group is
SU($N_c$).  We choose the quarks in Fig.\ \ref{fig:LO} to be in the
fundamental representation, and we choose the cross section to be
averaged over the colors of the initial-state quarks.  In a correct
factorization formula, the coefficient is independent of the nature of
the beam particles, of course, so these choices for the beams are
purely for calculational convenience.

Then the group theory factor for the graph for the cross section has
the form
\begin{equation}
  \frac{1}{N_c^2} \mbox{tr}( t_\alpha  t_{\alpha'} )
                  \mbox{tr}( t_\beta  t_{\beta'} )
                  f_{\alpha\beta\gamma}
                  f_{\alpha'\beta'\gamma}
   = \frac{ N_c^2-1 }{ 4 N_c }
.
\end{equation}
Here repeated indices are summed, as usual, and the result was
obtained by use of the formulas $\mbox{tr}(t_\alpha t_{\alpha'})=
\frac12 \delta_{\alpha\alpha'}$, $f_{\alpha\beta\gamma}
f_{\alpha\beta'\gamma}=N_c\delta_{\beta\beta'}$, and
$\delta_{\gamma\gamma}=N_c^2-1$.  

Each parton density has a group-theory factor
\begin{equation}
  \frac{1}{N_c} \mbox{tr}( t_\alpha  t_\alpha )
  = C_F
  = \frac{ N_c^2-1 }{ 2N_c }.
\end{equation}
The coefficient in the factorization formula therefore has a factor
\begin{equation}
  \frac{ N_c^2-1 }{ 4 N_c }
  \frac{1}{C_F^2}
 = \frac{ N_c }{ N_c^2 - 1 }
 = \frac{1}{ 2C_F },
\end{equation}
in agreement with our factorization formula in
\eqref{eq:ktfact.correct} and with the one in \cite{Kovchegov:2001sc},
but in disagreement with \cite{Gribov:1984tu}.

\sloppy
\bibliography{refs22}

\begin{thebibliography}{39}%
\makeatletter
\providecommand \@ifxundefined [1]{%
 \@ifx{#1\undefined}
}%
\providecommand \@ifnum [1]{%
 \ifnum #1\expandafter \@firstoftwo
 \else \expandafter \@secondoftwo
 \fi
}%
\providecommand \@ifx [1]{%
 \ifx #1\expandafter \@firstoftwo
 \else \expandafter \@secondoftwo
 \fi
}%
\providecommand \natexlab [1]{#1}%
\providecommand \enquote  [1]{``#1''}%
\providecommand \bibnamefont  [1]{#1}%
\providecommand \bibfnamefont [1]{#1}%
\providecommand \citenamefont [1]{#1}%
\providecommand \href@noop [0]{\@secondoftwo}%
\providecommand \href [0]{\begingroup \@sanitize@url \@href}%
\providecommand \@href[1]{\@@startlink{#1}\@@href}%
\providecommand \@@href[1]{\endgroup#1\@@endlink}%
\providecommand \@sanitize@url [0]{\catcode `\\12\catcode `\$12\catcode
  `\&12\catcode `\#12\catcode `\^12\catcode `\_12\catcode `\%12\relax}%
\providecommand \@@startlink[1]{}%
\providecommand \@@endlink[0]{}%
\providecommand \url  [0]{\begingroup\@sanitize@url \@url }%
\providecommand \@url [1]{\endgroup\@href {#1}{\urlprefix }}%
\providecommand \urlprefix  [0]{URL }%
\providecommand \Eprint [0]{\href }%
\providecommand \doibase [0]{http://dx.doi.org/}%
\providecommand \selectlanguage [0]{\@gobble}%
\providecommand \bibinfo  [0]{\@secondoftwo}%
\providecommand \bibfield  [0]{\@secondoftwo}%
\providecommand \translation [1]{[#1]}%
\providecommand \BibitemOpen [0]{}%
\providecommand \bibitemStop [0]{}%
\providecommand \bibitemNoStop [0]{.\EOS\space}%
\providecommand \EOS [0]{\spacefactor3000\relax}%
\providecommand \BibitemShut  [1]{\csname bibitem#1\endcsname}%
\let\auto@bib@innerbib\@empty
\bibitem [{\citenamefont {Kraml}\ \emph {et~al.}(2012)\citenamefont {Kraml},
  \citenamefont {Allanach}, \citenamefont {Mangano}, \citenamefont {Prosper},
  \citenamefont {Sekmen} \emph {et~al.}}]{Kraml:2012sg}%
  \BibitemOpen
  \bibfield  {author} {\bibinfo {author} {\bibfnamefont {S.}~\bibnamefont
  {Kraml}}, \bibinfo {author} {\bibfnamefont {B.C.}\ \bibnamefont {Allanach}},
  \bibinfo {author} {\bibfnamefont {M.}~\bibnamefont {Mangano}}, \bibinfo
  {author} {\bibfnamefont {H.B.}\ \bibnamefont {Prosper}}, \bibinfo {author}
  {\bibfnamefont {S.}~\bibnamefont {Sekmen}},  \emph {et~al.},\ }\bibfield
  {title} {\enquote {\bibinfo {title} {Searches for new physics: {L}es
  {H}ouches recommendations for the presentation of {LHC} results},}\
  }\href@noop {} {\  (\bibinfo {year} {2012})},\ \Eprint
  {http://arxiv.org/abs/1203.2489} {arXiv:1203.2489 [hep-ph]} \BibitemShut
  {NoStop}%
\bibitem [{\citenamefont {Collins}(2011)}]{qcdbook}%
  \BibitemOpen
  \bibfield  {author} {\bibinfo {author} {\bibfnamefont {J.~C.}\ \bibnamefont
  {Collins}},\ }\href@noop {} {\emph {\bibinfo {title} {Foundations of
  Perturbative {QCD}}}}\ (\bibinfo  {publisher} {Cambridge University Press},\
  \bibinfo {address} {Cambridge, UK},\ \bibinfo {year} {2011})\BibitemShut
  {NoStop}%
\bibitem [{Note1()}]{Note1}%
  \BibitemOpen
  \bibinfo {note} {Mark Strikman in a personal communication remarked that it
  is an interesting question as to whether the literature on Regge theory meets
  these standards sufficiently well. But we do not wish to answer that
  particular question here.}\BibitemShut {Stop}%
\bibitem [{\citenamefont {Kharzeev}\ \emph {et~al.}(2003)\citenamefont
  {Kharzeev}, \citenamefont {Kovchegov},\ and\ \citenamefont
  {Tuchin}}]{Kharzeev:2003wz}%
  \BibitemOpen
  \bibfield  {author} {\bibinfo {author} {\bibfnamefont {Dmitri}\ \bibnamefont
  {Kharzeev}}, \bibinfo {author} {\bibfnamefont {Yuri~V.}\ \bibnamefont
  {Kovchegov}}, \ and\ \bibinfo {author} {\bibfnamefont {Kirill}\ \bibnamefont
  {Tuchin}},\ }\bibfield  {title} {\enquote {\bibinfo {title} {{C}ronin effect
  and high {$p_T$} suppression in {pA} collisions},}\ }\href {\doibase
  10.1103/PhysRevD.68.094013} {\bibfield  {journal} {\bibinfo  {journal}
  {Phys.Rev.}\ }\textbf {\bibinfo {volume} {D68}},\ \bibinfo {pages} {094013}
  (\bibinfo {year} {2003})},\ \Eprint {http://arxiv.org/abs/hep-ph/0307037}
  {arXiv:hep-ph/0307037 [hep-ph]} \BibitemShut {NoStop}%
\bibitem [{\citenamefont {Armesto}\ \emph {et~al.}(2005)\citenamefont
  {Armesto}, \citenamefont {Salgado},\ and\ \citenamefont
  {Wiedemann}}]{Armesto:2004ud}%
  \BibitemOpen
  \bibfield  {author} {\bibinfo {author} {\bibfnamefont {Nestor}\ \bibnamefont
  {Armesto}}, \bibinfo {author} {\bibfnamefont {Carlos~A.}\ \bibnamefont
  {Salgado}}, \ and\ \bibinfo {author} {\bibfnamefont {Urs~Achim}\ \bibnamefont
  {Wiedemann}},\ }\bibfield  {title} {\enquote {\bibinfo {title} {Relating
  high-energy lepton-hadron, proton-nucleus and nucleus-nucleus collisions
  through geometric scaling},}\ }\href {\doibase 10.1103/PhysRevLett.94.022002}
  {\bibfield  {journal} {\bibinfo  {journal} {Phys.Rev.Lett.}\ }\textbf
  {\bibinfo {volume} {94}},\ \bibinfo {pages} {022002} (\bibinfo {year}
  {2005})},\ \Eprint {http://arxiv.org/abs/hep-ph/0407018}
  {arXiv:hep-ph/0407018 [hep-ph]} \BibitemShut {NoStop}%
\bibitem [{\citenamefont {Kharzeev}\ \emph {et~al.}(2005)\citenamefont
  {Kharzeev}, \citenamefont {Levin},\ and\ \citenamefont
  {Nardi}}]{Kharzeev:2004if}%
  \BibitemOpen
  \bibfield  {author} {\bibinfo {author} {\bibfnamefont {Dmitri}\ \bibnamefont
  {Kharzeev}}, \bibinfo {author} {\bibfnamefont {Eugene}\ \bibnamefont
  {Levin}}, \ and\ \bibinfo {author} {\bibfnamefont {Marzia}\ \bibnamefont
  {Nardi}},\ }\bibfield  {title} {\enquote {\bibinfo {title} {Color glass
  condensate at the {LHC}: {H}adron multiplicities in {$pp$}, {$pA$} and {$AA$}
  collisions},}\ }\href {\doibase 10.1016/j.nuclphysa.2004.10.018} {\bibfield
  {journal} {\bibinfo  {journal} {Nucl.Phys.}\ }\textbf {\bibinfo {volume}
  {A747}},\ \bibinfo {pages} {609--629} (\bibinfo {year} {2005})},\ \Eprint
  {http://arxiv.org/abs/hep-ph/0408050} {arXiv:hep-ph/0408050 [hep-ph]}
  \BibitemShut {NoStop}%
\bibitem [{\citenamefont {Levin}\ and\ \citenamefont
  {Rezaeian}(2010{\natexlab{a}})}]{Levin:2010dw}%
  \BibitemOpen
  \bibfield  {author} {\bibinfo {author} {\bibfnamefont {Eugene}\ \bibnamefont
  {Levin}}\ and\ \bibinfo {author} {\bibfnamefont {Amir~H.}\ \bibnamefont
  {Rezaeian}},\ }\bibfield  {title} {\enquote {\bibinfo {title} {Gluon
  saturation and inclusive hadron production at {LHC}},}\ }\href {\doibase
  10.1103/PhysRevD.82.014022} {\bibfield  {journal} {\bibinfo  {journal} {Phys.
  Rev.}\ }\textbf {\bibinfo {volume} {D82}},\ \bibinfo {pages} {014022}
  (\bibinfo {year} {2010}{\natexlab{a}})},\ \Eprint
  {http://arxiv.org/abs/1005.0631} {arXiv:1005.0631 [hep-ph]} \BibitemShut
  {NoStop}%
\bibitem [{\citenamefont {Levin}\ and\ \citenamefont
  {Rezaeian}(2010{\natexlab{b}})}]{Levin:2010zy}%
  \BibitemOpen
  \bibfield  {author} {\bibinfo {author} {\bibfnamefont {Eugene}\ \bibnamefont
  {Levin}}\ and\ \bibinfo {author} {\bibfnamefont {Amir~H.}\ \bibnamefont
  {Rezaeian}},\ }\bibfield  {title} {\enquote {\bibinfo {title} {Hadron
  multiplicity in {$pp$} and {$AA$} collisions at {LHC} from the color glass
  condensate},}\ }\href {\doibase 10.1103/PhysRevD.82.054003} {\bibfield
  {journal} {\bibinfo  {journal} {Phys. Rev.}\ }\textbf {\bibinfo {volume}
  {D82}},\ \bibinfo {pages} {054003} (\bibinfo {year} {2010}{\natexlab{b}})},\
  \Eprint {http://arxiv.org/abs/1007.2430} {arXiv:1007.2430 [hep-ph]}
  \BibitemShut {NoStop}%
\bibitem [{\citenamefont {Albacete}\ and\ \citenamefont
  {Marquet}(2010)}]{Albacete:2010bs}%
  \BibitemOpen
  \bibfield  {author} {\bibinfo {author} {\bibfnamefont {Javier~L.}\
  \bibnamefont {Albacete}}\ and\ \bibinfo {author} {\bibfnamefont {Cyrille}\
  \bibnamefont {Marquet}},\ }\bibfield  {title} {\enquote {\bibinfo {title}
  {Single inclusive hadron production at {RHIC} and the {LHC} from the color
  glass condensate},}\ }\href {\doibase 10.1016/j.physletb.2010.02.073}
  {\bibfield  {journal} {\bibinfo  {journal} {Phys.Lett.}\ }\textbf {\bibinfo
  {volume} {B687}},\ \bibinfo {pages} {174--179} (\bibinfo {year} {2010})},\
  \Eprint {http://arxiv.org/abs/1001.1378} {arXiv:1001.1378 [hep-ph]}
  \BibitemShut {NoStop}%
\bibitem [{\citenamefont {Albacete}\ and\ \citenamefont
  {Dumitru}(2010)}]{Albacete:2010ad}%
  \BibitemOpen
  \bibfield  {author} {\bibinfo {author} {\bibfnamefont {Javier~L.}\
  \bibnamefont {Albacete}}\ and\ \bibinfo {author} {\bibfnamefont {Adrian}\
  \bibnamefont {Dumitru}},\ }\bibfield  {title} {\enquote {\bibinfo {title} {A
  model for gluon production in heavy-ion collisions at the {LHC} with {rcBK}
  unintegrated gluon densities},}\ }\href@noop {} {\  (\bibinfo {year}
  {2010})},\ \Eprint {http://arxiv.org/abs/1011.5161} {arXiv:1011.5161
  [hep-ph]} \BibitemShut {NoStop}%
\bibitem [{\citenamefont {Rezaeian}(2012)}]{Rezaeian:2011ia}%
  \BibitemOpen
  \bibfield  {author} {\bibinfo {author} {\bibfnamefont {Amir~H.}\ \bibnamefont
  {Rezaeian}},\ }\bibfield  {title} {\enquote {\bibinfo {title} {Charged
  particle multiplicities in {pA} interactions at the {LHC} from the color
  glass condensate},}\ }\href@noop {} {\bibfield  {journal} {\bibinfo
  {journal} {Phys.Rev.}\ }\textbf {\bibinfo {volume} {D85}},\ \bibinfo {pages}
  {014028} (\bibinfo {year} {2012})},\ \Eprint {http://arxiv.org/abs/1111.2312}
  {arXiv:1111.2312 [hep-ph]} \BibitemShut {NoStop}%
\bibitem [{\citenamefont {Tribedy}\ and\ \citenamefont
  {Venugopalan}(2011)}]{Tribedy:2011aa}%
  \BibitemOpen
  \bibfield  {author} {\bibinfo {author} {\bibfnamefont {Prithwish}\
  \bibnamefont {Tribedy}}\ and\ \bibinfo {author} {\bibfnamefont {Raju}\
  \bibnamefont {Venugopalan}},\ }\bibfield  {title} {\enquote {\bibinfo {title}
  {{QCD} saturation at the {LHC}: comparisons of models to {p+p} and {A+A} data
  and predictions for {p+Pb} collisions},}\ }\href@noop {} {\  (\bibinfo {year}
  {2011})},\ \Eprint {http://arxiv.org/abs/1112.2445} {arXiv:1112.2445
  [hep-ph]} \BibitemShut {NoStop}%
\bibitem [{\citenamefont {Abelev}\ \emph {et~al.}(2010)\citenamefont {Abelev}
  \emph {et~al.}}]{Aamodt:2010pb}%
  \BibitemOpen
  \bibfield  {author} {\bibinfo {author} {\bibfnamefont {B}~\bibnamefont
  {Abelev}} \emph {et~al.} (\bibinfo {collaboration} {ALICE Collaboration}),\
  }\bibfield  {title} {\enquote {\bibinfo {title} {Charged-particle
  multiplicity density at mid-rapidity in central {Pb-Pb} collisions at
  {$\sqrt{s_{NN}} = 2.76$ TeV}},}\ }\href {\doibase
  10.1103/PhysRevLett.105.252301} {\bibfield  {journal} {\bibinfo  {journal}
  {Phys.Rev.Lett.}\ }\textbf {\bibinfo {volume} {105}},\ \bibinfo {pages}
  {252301} (\bibinfo {year} {2010})},\ \Eprint {http://arxiv.org/abs/1011.3916}
  {arXiv:1011.3916 [nucl-ex]} \BibitemShut {NoStop}%
\bibitem [{\citenamefont {Kovchegov}\ and\ \citenamefont
  {Tuchin}(2002)}]{Kovchegov:2001sc}%
  \BibitemOpen
  \bibfield  {author} {\bibinfo {author} {\bibfnamefont {Yuri~V.}\ \bibnamefont
  {Kovchegov}}\ and\ \bibinfo {author} {\bibfnamefont {Kirill}\ \bibnamefont
  {Tuchin}},\ }\bibfield  {title} {\enquote {\bibinfo {title} {Inclusive gluon
  production in {DIS} at high parton density},}\ }\href {\doibase
  10.1103/PhysRevD.65.074026} {\bibfield  {journal} {\bibinfo  {journal} {Phys.
  Rev.}\ }\textbf {\bibinfo {volume} {D65}},\ \bibinfo {pages} {074026}
  (\bibinfo {year} {2002})},\ \Eprint {http://arxiv.org/abs/hep-ph/0111362}
  {arXiv:hep-ph/0111362} \BibitemShut {NoStop}%
\bibitem [{\citenamefont {Gribov}\ \emph
  {et~al.}(1983{\natexlab{a}})\citenamefont {Gribov}, \citenamefont {Levin},\
  and\ \citenamefont {Ryskin}}]{Gribov:1984tu}%
  \BibitemOpen
  \bibfield  {author} {\bibinfo {author} {\bibfnamefont {L.~V.}\ \bibnamefont
  {Gribov}}, \bibinfo {author} {\bibfnamefont {E.~M.}\ \bibnamefont {Levin}}, \
  and\ \bibinfo {author} {\bibfnamefont {M.~G.}\ \bibnamefont {Ryskin}},\
  }\bibfield  {title} {\enquote {\bibinfo {title} {Semihard processes in
  {QCD}},}\ }\href {\doibase 10.1016/0370-1573(83)90022-4} {\bibfield
  {journal} {\bibinfo  {journal} {Phys. Rept.}\ }\textbf {\bibinfo {volume}
  {100}},\ \bibinfo {pages} {1--150} (\bibinfo {year}
  {1983}{\natexlab{a}})}\BibitemShut {NoStop}%
\bibitem [{\citenamefont {Kovner}\ \emph
  {et~al.}(1995{\natexlab{a}})\citenamefont {Kovner}, \citenamefont
  {McLerran},\ and\ \citenamefont {Weigert}}]{Kovner:1995ja}%
  \BibitemOpen
  \bibfield  {author} {\bibinfo {author} {\bibfnamefont {Alex}\ \bibnamefont
  {Kovner}}, \bibinfo {author} {\bibfnamefont {Larry~D.}\ \bibnamefont
  {McLerran}}, \ and\ \bibinfo {author} {\bibfnamefont {Heribert}\ \bibnamefont
  {Weigert}},\ }\bibfield  {title} {\enquote {\bibinfo {title} {Gluon
  production from non-abelian {W}eizsacker-{W}illiams fields in nucleus-nucleus
  collisions},}\ }\href {\doibase 10.1103/PhysRevD.52.6231} {\bibfield
  {journal} {\bibinfo  {journal} {Phys.Rev.}\ }\textbf {\bibinfo {volume}
  {D52}},\ \bibinfo {pages} {6231--6237} (\bibinfo {year}
  {1995}{\natexlab{a}})},\ \Eprint {http://arxiv.org/abs/hep-ph/9502289}
  {arXiv:hep-ph/9502289 [hep-ph]} \BibitemShut {NoStop}%
\bibitem [{\citenamefont {Kovner}\ \emph
  {et~al.}(1995{\natexlab{b}})\citenamefont {Kovner}, \citenamefont
  {McLerran},\ and\ \citenamefont {Weigert}}]{Kovner:1995ts}%
  \BibitemOpen
  \bibfield  {author} {\bibinfo {author} {\bibfnamefont {Alex}\ \bibnamefont
  {Kovner}}, \bibinfo {author} {\bibfnamefont {Larry~D.}\ \bibnamefont
  {McLerran}}, \ and\ \bibinfo {author} {\bibfnamefont {Heribert}\ \bibnamefont
  {Weigert}},\ }\bibfield  {title} {\enquote {\bibinfo {title} {Gluon
  production at high transverse momentum in the {M}c{L}erran-{V}enugopalan
  model of nuclear structure functions},}\ }\href {\doibase
  10.1103/PhysRevD.52.3809} {\bibfield  {journal} {\bibinfo  {journal}
  {Phys.Rev.}\ }\textbf {\bibinfo {volume} {D52}},\ \bibinfo {pages}
  {3809--3814} (\bibinfo {year} {1995}{\natexlab{b}})},\ \Eprint
  {http://arxiv.org/abs/hep-ph/9505320} {arXiv:hep-ph/9505320 [hep-ph]}
  \BibitemShut {NoStop}%
\bibitem [{\citenamefont {Kovchegov}\ and\ \citenamefont
  {Rischke}(1997)}]{Kovchegov:1997ke}%
  \BibitemOpen
  \bibfield  {author} {\bibinfo {author} {\bibfnamefont {Yuri~V.}\ \bibnamefont
  {Kovchegov}}\ and\ \bibinfo {author} {\bibfnamefont {Dirk~H.}\ \bibnamefont
  {Rischke}},\ }\bibfield  {title} {\enquote {\bibinfo {title} {Classical gluon
  radiation in ultrarelativistic nucleus-nucleus collisions},}\ }\href
  {\doibase 10.1103/PhysRevC.56.1084} {\bibfield  {journal} {\bibinfo
  {journal} {Phys.Rev.}\ }\textbf {\bibinfo {volume} {C56}},\ \bibinfo {pages}
  {1084--1094} (\bibinfo {year} {1997})},\ \Eprint
  {http://arxiv.org/abs/hep-ph/9704201} {arXiv:hep-ph/9704201 [hep-ph]}
  \BibitemShut {NoStop}%
\bibitem [{\citenamefont {Guo}(1999)}]{Guo:1998pe}%
  \BibitemOpen
  \bibfield  {author} {\bibinfo {author} {\bibfnamefont {Xiao-feng}\
  \bibnamefont {Guo}},\ }\bibfield  {title} {\enquote {\bibinfo {title} {Gluon
  minijet production in nuclear collisions at high-energies},}\ }\href
  {\doibase 10.1103/PhysRevD.59.094017} {\bibfield  {journal} {\bibinfo
  {journal} {Phys.Rev.}\ }\textbf {\bibinfo {volume} {D59}},\ \bibinfo {pages}
  {094017} (\bibinfo {year} {1999})},\ \Eprint
  {http://arxiv.org/abs/hep-ph/9812257} {arXiv:hep-ph/9812257 [hep-ph]}
  \BibitemShut {NoStop}%
\bibitem [{\citenamefont {Gyulassy}\ and\ \citenamefont
  {McLerran}(1997)}]{Gyulassy:1997vt}%
  \BibitemOpen
  \bibfield  {author} {\bibinfo {author} {\bibfnamefont {M.}~\bibnamefont
  {Gyulassy}}\ and\ \bibinfo {author} {\bibfnamefont {Larry~D.}\ \bibnamefont
  {McLerran}},\ }\bibfield  {title} {\enquote {\bibinfo {title} {{Y}ang-{M}ills
  radiation in ultrarelativistic nuclear collisions},}\ }\href {\doibase
  10.1103/PhysRevC.56.2219} {\bibfield  {journal} {\bibinfo  {journal}
  {Phys.Rev.}\ }\textbf {\bibinfo {volume} {C56}},\ \bibinfo {pages}
  {2219--2228} (\bibinfo {year} {1997})},\ \Eprint
  {http://arxiv.org/abs/nucl-th/9704034} {arXiv:nucl-th/9704034 [nucl-th]}
  \BibitemShut {NoStop}%
\bibitem [{\citenamefont {Avsar}(2012)}]{Avsar:2012hj}%
  \BibitemOpen
  \bibfield  {author} {\bibinfo {author} {\bibfnamefont {Emil}\ \bibnamefont
  {Avsar}},\ }\bibfield  {title} {\enquote {\bibinfo {title} {{TMD}
  factorization and the gluon distribution in high energy {QCD}},}\ }\href@noop
  {} {\  (\bibinfo {year} {2012})},\ \Eprint {http://arxiv.org/abs/1203.1916}
  {arXiv:1203.1916 [hep-ph]} \BibitemShut {NoStop}%
\bibitem [{\citenamefont {Gribov}\ \emph
  {et~al.}(1981{\natexlab{a}})\citenamefont {Gribov}, \citenamefont {Levin},\
  and\ \citenamefont {Ryskin}}]{Gribov:1981kg}%
  \BibitemOpen
  \bibfield  {author} {\bibinfo {author} {\bibfnamefont {L.~V.}\ \bibnamefont
  {Gribov}}, \bibinfo {author} {\bibfnamefont {E.~M.}\ \bibnamefont {Levin}}, \
  and\ \bibinfo {author} {\bibfnamefont {M.~G.}\ \bibnamefont {Ryskin}},\
  }\bibfield  {title} {\enquote {\bibinfo {title} {High $p_t$ hadrons in the
  pionization region in {QCD}},}\ }\href {\doibase
  10.1016/0370-2693(81)90768-1} {\bibfield  {journal} {\bibinfo  {journal}
  {Phys. Lett.}\ }\textbf {\bibinfo {volume} {B100}},\ \bibinfo {pages}
  {173--176} (\bibinfo {year} {1981}{\natexlab{a}})}\BibitemShut {NoStop}%
\bibitem [{\citenamefont {Gribov}\ \emph
  {et~al.}(1983{\natexlab{b}})\citenamefont {Gribov}, \citenamefont {Levin},\
  and\ \citenamefont {Ryskin}}]{Gribov:1983fc}%
  \BibitemOpen
  \bibfield  {author} {\bibinfo {author} {\bibfnamefont {L.~V.}\ \bibnamefont
  {Gribov}}, \bibinfo {author} {\bibfnamefont {E.~M.}\ \bibnamefont {Levin}}, \
  and\ \bibinfo {author} {\bibfnamefont {M.~G.}\ \bibnamefont {Ryskin}},\
  }\bibfield  {title} {\enquote {\bibinfo {title} {Large-{$E_T$} processes as a
  main source of hadrons at very high-energies},}\ }\href {\doibase
  10.1016/0370-2693(83)90204-6} {\bibfield  {journal} {\bibinfo  {journal}
  {Phys. Lett.}\ }\textbf {\bibinfo {volume} {B121}},\ \bibinfo {pages}
  {65--71} (\bibinfo {year} {1983}{\natexlab{b}})}\BibitemShut {NoStop}%
\bibitem [{\citenamefont {Ryskin}(1980)}]{Ryskin:1980yz}%
  \BibitemOpen
  \bibfield  {author} {\bibinfo {author} {\bibfnamefont {M.G.}\ \bibnamefont
  {Ryskin}},\ }\bibfield  {title} {\enquote {\bibinfo {title} {Hadron productin
  with large transverse momentum in the pionization region and the vacuum
  singularity in {QCD}},}\ }\href@noop {} {\bibfield  {journal} {\bibinfo
  {journal} {Sov. J. Nucl. Phys.}\ }\textbf {\bibinfo {volume} {32}},\ \bibinfo
  {pages} {133--140} (\bibinfo {year} {1980})},\ \bibinfo {note} {yad. Fiz. 32,
  259-274 (1980)}\BibitemShut {NoStop}%
\bibitem [{\citenamefont {Gribov}\ \emph
  {et~al.}(1981{\natexlab{b}})\citenamefont {Gribov}, \citenamefont {Levin},\
  and\ \citenamefont {Ryskin}}]{Gribov:1980uj}%
  \BibitemOpen
  \bibfield  {author} {\bibinfo {author} {\bibfnamefont {L.V.}\ \bibnamefont
  {Gribov}}, \bibinfo {author} {\bibfnamefont {E.M.}\ \bibnamefont {Levin}}, \
  and\ \bibinfo {author} {\bibfnamefont {M.G.}\ \bibnamefont {Ryskin}},\
  }\bibfield  {title} {\enquote {\bibinfo {title} {Extending the {QCD} leading
  logs up to reggeon field theory},}\ }\href {\doibase
  10.1016/0370-2693(81)90669-9} {\bibfield  {journal} {\bibinfo  {journal}
  {Phys.Lett.}\ }\textbf {\bibinfo {volume} {B101}},\ \bibinfo {pages} {185}
  (\bibinfo {year} {1981}{\natexlab{b}})}\BibitemShut {NoStop}%
\bibitem [{\citenamefont {Collins}\ and\ \citenamefont
  {Qiu}(2007)}]{Collins:2007nk}%
  \BibitemOpen
  \bibfield  {author} {\bibinfo {author} {\bibfnamefont {John}\ \bibnamefont
  {Collins}}\ and\ \bibinfo {author} {\bibfnamefont {Jian-Wei}\ \bibnamefont
  {Qiu}},\ }\bibfield  {title} {\enquote {\bibinfo {title} {{$k_{T}$}
  factorization is violated in production of high-transverse-momentum particles
  in hadron-hadron collisions},}\ }\href {\doibase 10.1103/PhysRevD.75.114014}
  {\bibfield  {journal} {\bibinfo  {journal} {Phys. Rev.}\ }\textbf {\bibinfo
  {volume} {D75}},\ \bibinfo {pages} {114014} (\bibinfo {year} {2007})},\
  \Eprint {http://arxiv.org/abs/0705.2141} {arXiv:0705.2141 [hep-ph]}
  \BibitemShut {NoStop}%
\bibitem [{\citenamefont {Rogers}\ and\ \citenamefont
  {Mulders}(2010)}]{Rogers:2010dm}%
  \BibitemOpen
  \bibfield  {author} {\bibinfo {author} {\bibfnamefont {Ted~C.}\ \bibnamefont
  {Rogers}}\ and\ \bibinfo {author} {\bibfnamefont {Piet~J.}\ \bibnamefont
  {Mulders}},\ }\bibfield  {title} {\enquote {\bibinfo {title} {No generalized
  {TMD}-factorization in hadro-production of high transverse momentum
  hadrons},}\ }\href {\doibase 10.1103/PhysRevD.81.094006} {\bibfield
  {journal} {\bibinfo  {journal} {Phys.Rev.}\ }\textbf {\bibinfo {volume}
  {D81}},\ \bibinfo {pages} {094006} (\bibinfo {year} {2010})},\ \Eprint
  {http://arxiv.org/abs/1001.2977} {arXiv:1001.2977 [hep-ph]} \BibitemShut
  {NoStop}%
\bibitem [{\citenamefont {Forshaw}\ \emph {et~al.}(2012)\citenamefont
  {Forshaw}, \citenamefont {Seymour},\ and\ \citenamefont
  {Siodmok}}]{Forshaw:2012bi}%
  \BibitemOpen
  \bibfield  {author} {\bibinfo {author} {\bibfnamefont {Jeffrey~R.}\
  \bibnamefont {Forshaw}}, \bibinfo {author} {\bibfnamefont {Michael~H.}\
  \bibnamefont {Seymour}}, \ and\ \bibinfo {author} {\bibfnamefont {Andrzej}\
  \bibnamefont {Siodmok}},\ }\bibfield  {title} {\enquote {\bibinfo {title} {On
  the breaking of collinear factorization in {QCD}},}\ }\href@noop {} {\
  (\bibinfo {year} {2012})},\ \Eprint {http://arxiv.org/abs/1206.6363}
  {arXiv:1206.6363 [hep-ph]} \BibitemShut {NoStop}%
\bibitem [{Note2()}]{Note2}%
  \BibitemOpen
  \bibinfo {note} {\label {fn:mismatch1}In addition, the foundational work on a
  topic, which may be decades old, may give the key ideas, but not the extra
  steps to put results in the form in which they are later found to be most
  useful and convenient. It may be that these extra steps are known privately
  to insiders but that they do not appear in the papers that it is most natural
  to cite. We give an explicit example in Sec.\ \ref {sec:gluon.density.def}.
  This phenomenon can explain the problems with the citations, but it leaves
  unchanged the readers' difficulty of reproducing results.}\BibitemShut
  {Stop}%
\bibitem [{\citenamefont {Collins}\ and\ \citenamefont
  {Soper}(1982)}]{Collins:1981uw}%
  \BibitemOpen
  \bibfield  {author} {\bibinfo {author} {\bibfnamefont {John~C.}\ \bibnamefont
  {Collins}}\ and\ \bibinfo {author} {\bibfnamefont {Davison~E.}\ \bibnamefont
  {Soper}},\ }\bibfield  {title} {\enquote {\bibinfo {title} {Parton
  distribution and decay functions},}\ }\href {\doibase
  10.1016/0550-3213(82)90021-9} {\bibfield  {journal} {\bibinfo  {journal}
  {Nucl. Phys.}\ }\textbf {\bibinfo {volume} {B194}},\ \bibinfo {pages} {445}
  (\bibinfo {year} {1982})}\BibitemShut {NoStop}%
\bibitem [{\citenamefont {Collins}\ and\ \citenamefont
  {Soper}(1981)}]{Collins:1981uk}%
  \BibitemOpen
  \bibfield  {author} {\bibinfo {author} {\bibfnamefont {John~C.}\ \bibnamefont
  {Collins}}\ and\ \bibinfo {author} {\bibfnamefont {Davison~E.}\ \bibnamefont
  {Soper}},\ }\bibfield  {title} {\enquote {\bibinfo {title} {Back-to-back jets
  in {QCD}},}\ }\href {\doibase 10.1016/0550-3213(81)90339-4} {\bibfield
  {journal} {\bibinfo  {journal} {Nucl. Phys.}\ }\textbf {\bibinfo {volume}
  {B193}},\ \bibinfo {pages} {381} (\bibinfo {year} {1981})}\BibitemShut
  {NoStop}%
\bibitem [{\citenamefont {Bouchiat}\ \emph {et~al.}(1971)\citenamefont
  {Bouchiat}, \citenamefont {Fayet},\ and\ \citenamefont
  {Meyer}}]{Bouchiat:1971mj}%
  \BibitemOpen
  \bibfield  {author} {\bibinfo {author} {\bibfnamefont {C.}~\bibnamefont
  {Bouchiat}}, \bibinfo {author} {\bibfnamefont {P.}~\bibnamefont {Fayet}}, \
  and\ \bibinfo {author} {\bibfnamefont {P.}~\bibnamefont {Meyer}},\ }\bibfield
   {title} {\enquote {\bibinfo {title} {Galilean invariance in the infinite
  momentum frame and the parton model},}\ }\href@noop {} {\bibfield  {journal}
  {\bibinfo  {journal} {Nucl. Phys.}\ }\textbf {\bibinfo {volume} {B34}},\
  \bibinfo {pages} {157--176} (\bibinfo {year} {1971})}\BibitemShut {NoStop}%
\bibitem [{\citenamefont {Soper}(1977)}]{Soper:1977jc}%
  \BibitemOpen
  \bibfield  {author} {\bibinfo {author} {\bibfnamefont {D.~E.}\ \bibnamefont
  {Soper}},\ }\bibfield  {title} {\enquote {\bibinfo {title} {The parton model
  and the {B}ethe-{S}alpeter wave function},}\ }\href@noop {} {\bibfield
  {journal} {\bibinfo  {journal} {Phys. Rev.}\ }\textbf {\bibinfo {volume}
  {D15}},\ \bibinfo {pages} {1141--1149} (\bibinfo {year} {1977})}\BibitemShut
  {NoStop}%
\bibitem [{Note3()}]{Note3}%
  \BibitemOpen
  \bibinfo {note} {This gives an example of the difficulty referred to in
  footnote \ref {fn:mismatch1}}\BibitemShut {NoStop}%
\bibitem [{\citenamefont {Soper}(1979)}]{Soper:1979fq}%
  \BibitemOpen
  \bibfield  {author} {\bibinfo {author} {\bibfnamefont {D.~E.}\ \bibnamefont
  {Soper}},\ }\bibfield  {title} {\enquote {\bibinfo {title} {Partons and their
  transverse momenta in {QCD}},}\ }\href@noop {} {\bibfield  {journal}
  {\bibinfo  {journal} {Phys. Rev. Lett.}\ }\textbf {\bibinfo {volume} {43}},\
  \bibinfo {pages} {1847--1851} (\bibinfo {year} {1979})}\BibitemShut {NoStop}%
\bibitem [{Note4()}]{Note4}%
  \BibitemOpen
  \bibinfo {note} {Note that the definition in \cite {qcdbook} is for a TMD
  quark density. For a TMD gluon density, one makes the same adjustments as are
  used to get the formula for an integrated gluon density from the formula for
  an integrated quark density.}\BibitemShut {Stop}%
\bibitem [{\citenamefont {Bacchetta}\ \emph {et~al.}(2008)\citenamefont
  {Bacchetta}, \citenamefont {Boer}, \citenamefont {Diehl},\ and\ \citenamefont
  {Mulders}}]{Bacchetta:2008xw}%
  \BibitemOpen
  \bibfield  {author} {\bibinfo {author} {\bibfnamefont {Alessandro}\
  \bibnamefont {Bacchetta}}, \bibinfo {author} {\bibfnamefont {Daniel}\
  \bibnamefont {Boer}}, \bibinfo {author} {\bibfnamefont {Markus}\ \bibnamefont
  {Diehl}}, \ and\ \bibinfo {author} {\bibfnamefont {Piet~J.}\ \bibnamefont
  {Mulders}},\ }\bibfield  {title} {\enquote {\bibinfo {title} {Matches and
  mismatches in the descriptions of semi-inclusive processes at low and high
  transverse momentum},}\ }\href {\doibase 10.1088/1126-6708/2008/08/023}
  {\bibfield  {journal} {\bibinfo  {journal} {JHEP}\ }\textbf {\bibinfo
  {volume} {08}},\ \bibinfo {pages} {023} (\bibinfo {year} {2008})},\ \Eprint
  {http://arxiv.org/abs/0803.0227} {arXiv:0803.0227 [hep-ph]} \BibitemShut
  {NoStop}%
\bibitem [{\citenamefont {Bodwin}(1985)}]{Bodwin:1985hc}%
  \BibitemOpen
  \bibfield  {author} {\bibinfo {author} {\bibfnamefont {G.~T.}\ \bibnamefont
  {Bodwin}},\ }\bibfield  {title} {\enquote {\bibinfo {title} {Factorization of
  the {D}rell-{Y}an cross-section in perturbation theory},}\ }\href@noop {}
  {\bibfield  {journal} {\bibinfo  {journal} {Phys. Rev.}\ }\textbf {\bibinfo
  {volume} {D31}},\ \bibinfo {pages} {2616--2642} (\bibinfo {year} {1985})},\
  \bibinfo {note} {erratum: D34, 3932 (1986)}\BibitemShut {NoStop}%
\bibitem [{\citenamefont {Collins}\ \emph {et~al.}(1988)\citenamefont
  {Collins}, \citenamefont {Soper},\ and\ \citenamefont
  {Sterman}}]{Collins:1988ig}%
  \BibitemOpen
  \bibfield  {author} {\bibinfo {author} {\bibfnamefont {J.~C.}\ \bibnamefont
  {Collins}}, \bibinfo {author} {\bibfnamefont {D.~E.}\ \bibnamefont {Soper}},
  \ and\ \bibinfo {author} {\bibfnamefont {G.}~\bibnamefont {Sterman}},\
  }\bibfield  {title} {\enquote {\bibinfo {title} {Soft gluons and
  factorization},}\ }\href@noop {} {\bibfield  {journal} {\bibinfo  {journal}
  {Nucl. Phys.}\ }\textbf {\bibinfo {volume} {B308}},\ \bibinfo {pages}
  {833--856} (\bibinfo {year} {1988})}\BibitemShut {NoStop}%
\end{thebibliography}%

\end{document}